\documentclass[twocolumn,showpacs,superscriptaddress,prb]{revtex4-1}
\usepackage{graphicx}
\usepackage{textcomp}
\usepackage{amsmath,amssymb,amsfonts,amsthm,latexsym} 
\usepackage[colorlinks=true,urlcolor=blue,linkcolor=blue,citecolor=blue,a4paper]{hyperref}
\usepackage{subfig}
\usepackage{gensymb}
\usepackage{setspace}

 \makeatletter
    \renewcommand\@make@capt@title[2]{%
     \@ifx@empty\float@link{\@firstofone}{\expandafter\href\expandafter{\float@link}}%
      {\textbf{#1}}\@caption@fignum@sep#2\quad}%
    \makeatother
\usepackage[labelsep=quad,justification=raggedright]{caption}

\usepackage{color}      

\begin{document}

\title{In-situ spectroscopy of intrinsic Bi$_2$Te$_3$ topological insulator thin films and impact of extrinsic defects}
\author{P. Ngabonziza}
\affiliation{Faculty of Science and Technology and MESA+ Institute for Nanotechnology, University of Twente, 7500 AE Enschede, The Netherlands}
 \author{R. Heimbuch}
 \affiliation{Faculty of Science and Technology and MESA+ Institute for Nanotechnology, University of Twente, 7500 AE Enschede, The Netherlands}
\author{N. de Jong }
\affiliation{Van der Waals-Zeeman Institute, University of Amsterdam, Science Park 904, 1098 XH, Amsterdam, Netherlands} 
\author{R. A. Klaassen}
\affiliation{Faculty of Science and Technology and MESA+ Institute for Nanotechnology, University of Twente, 7500 AE Enschede, The Netherlands}
\author{M. P. Stehno}
\affiliation{Faculty of Science and Technology and MESA+ Institute for Nanotechnology, University of Twente, 7500 AE Enschede, The Netherlands}
\author{M. Snelder}
\affiliation{Faculty of Science and Technology and MESA+ Institute for Nanotechnology, University of Twente, 7500 AE Enschede, The Netherlands}
\author{A. Solmaz}
\affiliation{Faculty of Science and Technology and MESA+ Institute for Nanotechnology, University of Twente, 7500 AE Enschede, The Netherlands}
\author{S. V. Ramankutty}
\affiliation{Van der Waals-Zeeman Institute, University of Amsterdam, Science Park 904, 1098 XH, Amsterdam, Netherlands}
\author{E. Frantzeskakis}
\affiliation{Van der Waals-Zeeman Institute, University of Amsterdam, Science Park 904, 1098 XH, Amsterdam, Netherlands}
\author{ E. van Heumen}
\affiliation{Van der Waals-Zeeman Institute, University of Amsterdam, Science Park 904, 1098 XH, Amsterdam, Netherlands}

\author{G. Koster}
\affiliation{Faculty of Science and Technology and MESA+ Institute for Nanotechnology, University of Twente, 7500 AE Enschede, The Netherlands}
\author{M. S. Golden}
\affiliation{Van der Waals-Zeeman Institute, University of Amsterdam, Science Park 904, 1098 XH, Amsterdam, Netherlands}
\author{H. J. W. Zandvliet}
\affiliation{Faculty of Science and Technology and MESA+ Institute for Nanotechnology, University of Twente, 7500 AE Enschede, The Netherlands}
\author{A. Brinkman}
\affiliation{Faculty of Science and Technology and MESA+ Institute for Nanotechnology, University of Twente, 7500 AE Enschede, The Netherlands}
\date{\today}

\begin{abstract}
Combined in-situ x-ray photoemission spectroscopy, scanning tunnelling spectroscopy and angle resolved photoemission spectroscopy of molecular beam epitaxy grown Bi$_2$Te$_3$ on lattice mismatched substrates reveal high quality stoichiometric thin films with topological surface states without a contribution from the bulk bands at the Fermi energy. The absence of bulk states at the Fermi energy is achieved without counter doping. We observe that the surface morphology and electronic band structure of Bi$_2$Te$_3$ are not affected by in-vacuo storage and exposure to oxygen, whereas major changes are observed when exposed to ambient conditions. These films help define a pathway towards intrinsic topological devices.
\end{abstract}

\pacs{79.60.Dp, 73.20.-r, 68.37.Ef, 79.60.-i}
\maketitle
Topological insulators (TIs) are materials with an insulating bulk interior and spin-momentum-locked metallic surface states as a result of a band inversion from large spin-orbit interaction \cite{Fu2007,Zhang2009}. Bismuth telluride (Bi$_2$Te$_3$) is one of the 3D TI materials that has received a considerable amount of  attention as a potential candidate for room temperature spintronics and quantum computational devices \cite{Moore2010}. However, despite significant progress in bulk preparation techniques of TI materials, growing high-quality bulk Bi$_2$Te$_3$ crystals with a low number of defects and without shunt conduction through the bulk of the material is still a major challenge. The bulk carrier conduction complicates the direct exploitation of the remarkable properties of TI surfaces.  

Molecular beam epitaxy (MBE) is an established method for growing high-quality crystalline thin films of TIs with surface-dominated conduction \cite{He2012,Taskin2012,He2013}. Using this technique, it has been possible to prepare thin films that are insulating in the bulk. By varying the growth parameters and using substrates with negligible lattice mismatch, bulk insulating thin films of Bi$_2$Te$_3$ have been synthesized\cite{Li2010,Wang2011,Hoefer2014}. It would be good, on the other hand, to investigate the growth mode of high quality intrinsic Bi$_2$Te$_3$ films on lattice mismatched substrates and especially on insulating substrates (with high relative dielectric constant), as these offer the prospect of strong, gate-induced modulation of the sample's carrier density\cite{XHe2012}.

One of the current challenges, though, for realizing TI devices, is that grown films are often exposed to air during mounting of contacts or subsequent device fabrication steps. This leads to possible shifts of the Fermi level which result in enhanced bulk conductance\cite{Benia2011,CChen2012}. To date, it is not well-understood how \textit{ex-situ} contamination processes affect surface and bulk states of Bi$_2$Te$_3$ films so that precautions can be taken before taking them\textit{ ex-situ} for further investigations. Recently, it was shown for bulk-insulating  Bi$_2$Te$_3$ films that pure oxygen exposure, at low pressure ($10^{-6}$ mbar) has no significant influence on their charge transport properties\cite{Hoefer2014}. Nevertheless, since most magnetotransport studies and fabrication processes are carried out in ambient conditions, it is good to understand the impact of oxygen at atmospheric pressure, exposure to air and other \textit{ex-situ} contaminations. For these investigations, it is necessary to have high quality films since their inertness to oxidation, for example, depends on the amount of surface defects and grain boundaries.

In this work, we first focus on optimizing our growth procedure in order to realize bulk insulating films grown on lattice mismatched\cite{Eibl2015} insulating substrates (Al$_2$O$_3$[0001] and SrTiO$_3$[111]). Secondly, we perform a systematic \textit{in-situ} characterization to investigate the effect of aging/degradation due to any vacancies or antisite defects for films stored in-vacuo. 
Thirdly, our \textit{in-situ} angle resolved photoemission spectroscopy (ARPES) experiments clearly reveal topological Dirac surface states, consistent with a linear increase in the density of states (DOS) measured by in-situ scanning tunneling spectroscopy. Lastly, we study the effect of pure oxygen exposure at atmospheric pressure; and then the effect of \textit{ex-situ} contamination in air. We use x-ray photoemission spectroscopy (XPS) for the characterization and detection of any Te/Bi excess in our films, and we employ low temperature scanning tunnelling spectroscopy (STS) to acquire differential conductivity spectra for samples kept \textit{in-situ}, then exposed to pure oxygen at atmospheric pressure and later to air. Our STS analysis is further confirmed by \textit{in-situ} ARPES investigations at low temperature and room temperature for samples kept \textit{in-situ} and later exposed to ambient conditions. 
\begin{figure*}[t]
{\includegraphics[width=1.\textwidth]{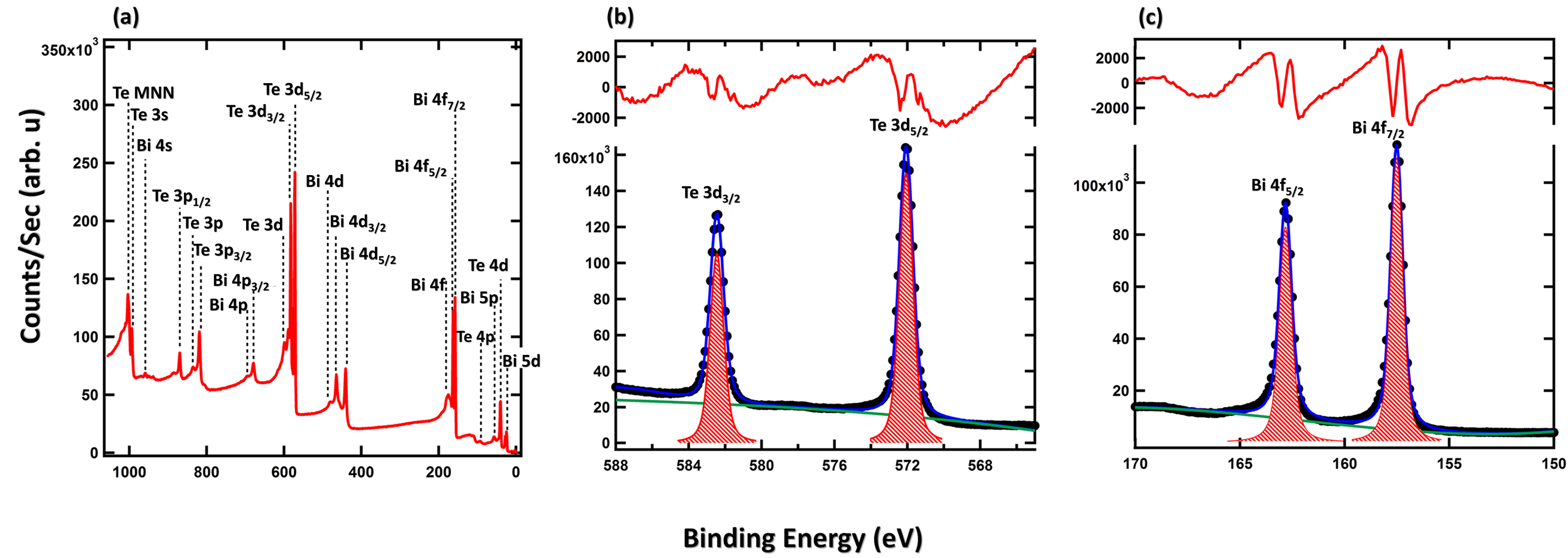}}  
  \caption{ \textit{In-situ} measurements of a 30 nm Bi$_2$Te$_3$ film grown on Al$_2$O$_3 [0001]$. (a) General scan. Only Bi and Te peaks are resolved. (b) and (c) High-resolution scans around the Te $3d$ and Bi $4f$ main peaks, respectively . The red shaded zone shows the areas of the fitted peaks with the background (green line) removed. The upper red curve is the residual after Shirley background subtraction.}
  \label{fig:XPS}
\end{figure*}

Our spectroscopy data (STS and ARPES) reveal that the as-grown Bi$_2$Te$_3$ films on lattice mismatched insulating substrates possess an intrinsically topological electronic structure, meaning that the Fermi level only crosses the topologically non-trivial metallic surface states. This is achieved without any counterdoping. Combining \textit{in-situ }and \textit{ex-situ} spectroscopy and topography data, we find that \textit{in-situ} storage in ultra-high vacuum and (short) exposure to pure oxygen at atmospheric pressures leave the band structure at the surface unaffected whereas breaking vacuum and exposing the sample surface to air results in notable changes in the surface band structure spectra and topography of the films.

Thin films of Bi$_2$Te$_3$ were grown on c-plane sapphire Al$_2$O$_3 [0001]$ and SrTiO$_3$[111] (STO) substrates  using MBE by co-evaporating high purity Te (99.999$\%$) and Bi (99.999$\%$) in a Te rich environment. The base pressure in the deposition chamber was lower than $5\times10^{-10}$ mbar. During deposition, the flux ratio Te/Bi was kept at about 10, the growth rate at $\sim 4 \textrm{ \AA}/\text{min}$ and the highest pressure recorded was $2.8\times10^{-8}$ mbar.
Before introducing the sapphire substrates in the MBE chamber, they were cleaned with acetone and ethanol in an ultrasonic bath, then annealed at 1050\degree C for 1 hour at atmospheric conditions. This cleaning procedure resulted in straight step edges and atomically flat terraces at the substrate surface, with a width of $\sim$ 150 nm. 
For the STO [111] substrates, to achieve atomically flat surfaces before loading them into the MBE chamber, we used a method similar to that reported in Refs. \cite{Chang2008,GZhang2011}. In order to avoid disordered interfacial layers at the interfaces between our films and the substrate, we used a two-step temperature growth scheme \cite{Bansal2011,Harrison2013}, which results in atomically sharp interfaces between the TI film and the substrate. The first nucleation layer was deposited at a temperature of 190\degree C, then slowly annealed to higher growth temperature of 230\degree C in order to improve the crystalline quality. The annealed layer was then used as a template for subsequent epitaxial growth of the second layer. Compared to films grown in a one-step temperature scheme, we observe a much better quality for the two-step growth procedure in terms of a flat morphology, and considerably reduced number of 3D growth defects. Similar significant reduction of 3D structure density has also been reported recently on two-step grown Bi$_2$Te$_3$ films\cite{Harrison2013}. To further improve the surface smoothness and quality of the films, we continued to anneal at 230\degree C for an additional 30 minutes; then cooled down to room-temperature at a rate of 3\degree /min. This growth procedure yields high quality Bi$_2$Te$_3$ films. Using this growth procedure, we present in this study spectroscopy data of two samples of 30 and 20 nm grown on Al$_2$O$_3 [0001]$ and one 15 nm film grown on SrTiO${_3}$ [111].

After growth, samples were characterized \textit{in-situ} using XPS to investigate the surface elemental composition and stoichiometry. Samples were transferred to XPS without breaking ultra-high vacuum (UHV) conditions since the XPS system is connected to the growth chamber via a distribution chamber. The UHV conditions of this system ensure that films are free from \textit{ex-situ} contamination. Figure \ref{fig:XPS}\textcolor{blue}{(a)} shows an \textit{in-situ} XPS scan, where only Bi and Te peaks are resolved with no appearance of extra peaks such as carbon and oxygen as often seen in films exposed to atmosphere \cite{Bando2000,Kong2011,Guo2013,Yashina2013} (see Fig. \textcolor{blue}{S2(a)} in the supplementary information). This lack of other core level signatures is an indication of a clean surface, free from contaminations. The surface chemical stoichiometry was studied by fitting the area under the Te $3d_{5/2}$, Te $3d_{3/2}$ and Bi $4f_{7/2}$, Bi $4f_{5/2}$ peaks. The ratio is determined to be $1.49\pm 0.05$. Figures  \ref{fig:XPS}\textcolor{blue}{(b)} and \textcolor{blue}{(c)} show these Te and Bi peaks, respectively, together with the Voigt function fits (after subtraction of Shirley background). 
The peak positions and their relative intensities are consistent with that of Bi$_2$Te$_3$ samples reported in the literature \cite{Bando2000,Zhang2004,Roy2013}.

To further confirm the high crystalline quality of our Bi$_2$Te$_3$ films after \textit{in-situ} measurements (STM/STS and ARPES), we performed \textit{ex-situ} x-ray diffraction (XRD) measurements, which confirmed that the films were grown with in-plane and out-of-plane lattice constants $a=  4.34\text{ \AA}$ and $c=30.41\text{ \AA}$, respectively. These lattice parameters are consistent with values reported previously for Bi$_2$Te$_3$ films \cite{Harrison2013,Park2012}. From XRD measurements of films grown on Al$_2$O$_3 [0001]$ and STO[111] substrates  (see Fig. \textcolor{blue}{S3} in the supplementary information), in the $2\theta-\omega$ scan only substrate peaks and the (0 0 3) of diffraction peaks from the films are resolved; implying the films are aligned along the c-axis. From the rocking curves, the measured full width half maximum of the (0 0 6) peak are 0.0414\degree (for films grown on sapphire) and 0.0417\degree    
(for films grown on STO), confirming the excellent quality of our films. Despite the lattice mismatch between Bi$_2$Te$_3$ [001] and  these substrates, smooth films had formed. This is because of the Van der Waals epitaxy\cite{He2013,Koma1992_1,Koma1992_2}, which relaxes the lattice-matching condition required for most common epitaxial growth of covalent semiconductors and their heterostructures.
\begin{figure}[t!]
{\includegraphics[width=0.5\textwidth]{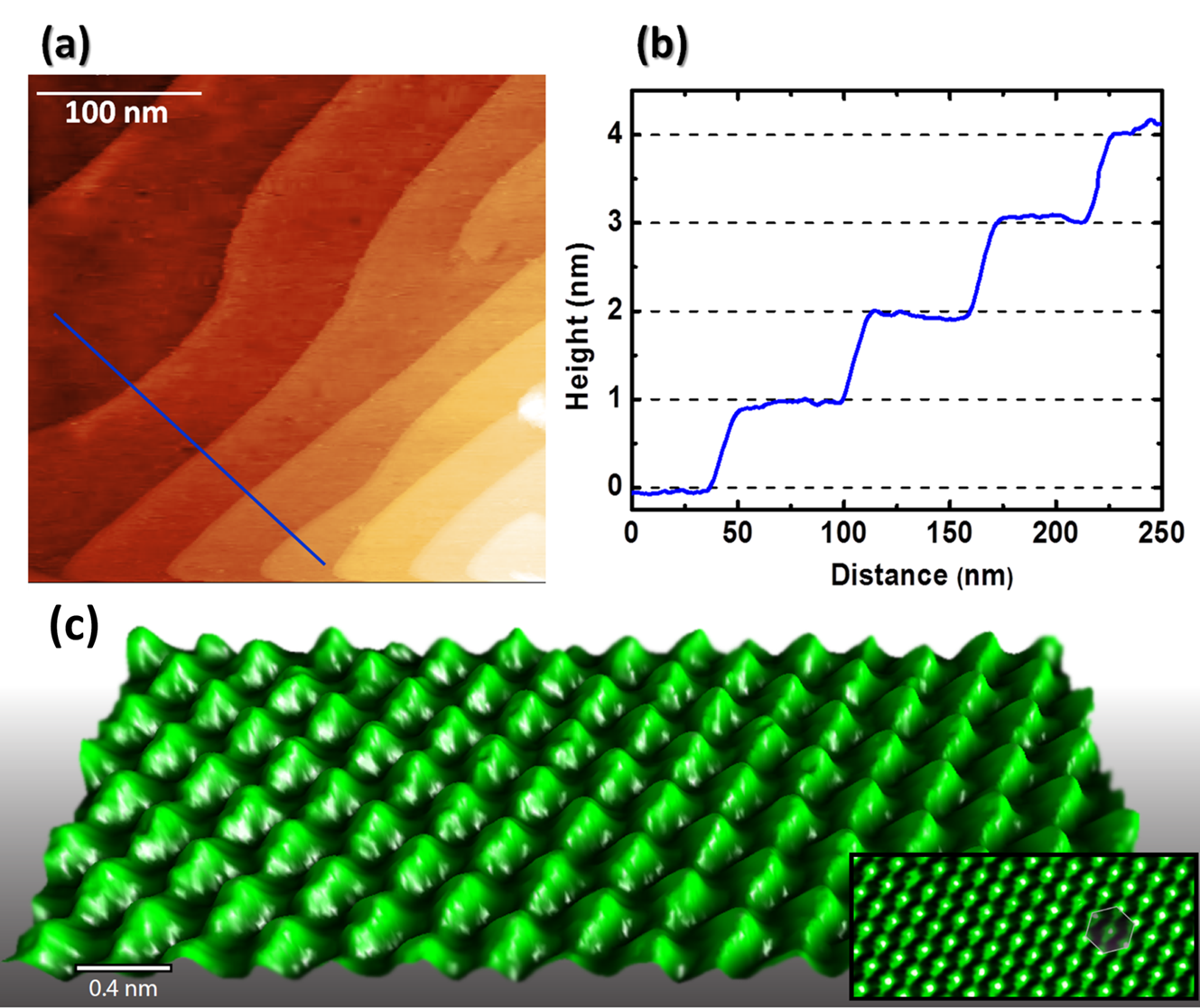}}         
  \caption{Topography of 30 nm Bi$_2$Te$_3$ film grown on sapphire. (a) A scan of a 250$\times$250 nm$^2$ area showing smooth film terraces. (b) Line profile across a series of wide steps (blue line). The step height is $\sim 1 \text{ nm}$. (c) Atomic resolution of a $6 \times 2.6$ nm$^2$ area showing the surface tellurium atoms with a hexagonal crystal structure (right bottom insert). The secondary lobes are artefacts caused by a slight doubling of the STM tip.}
  \label{fig:STM_topography}
\end{figure}

After \textit{in-situ} XPS measurements, samples were subsequently transferred to the low temperature STM chamber, which is not connected to the growth chamber. The transfer was done using a UHV suitcase equipped with a nonevaporable getter (NEG) pump \cite{Firpo2004} together with an ion pump. During the transfer process, the pressure remained lower than $2\times10^{-9}$ mbar.  Once samples were transferred to the STM, its UHV conditions (in the $10^{-12}$ mbar range) ensured that the film surface remained free from collecting further adsorbates for months. The \textit{in-situ} STM/STS data were acquired at 77 K. Many topography and spectroscopy maps were taken at random positions on the surface of the film at a bias voltage and set-point current of +350 mV and 1 nA. The results were consistent with previously published STM studies on Bi$_2$Te$_3$\cite{Urazhdin2004,Alpichshev2010,Chen2012}. Figure \ref{fig:STM_topography}\textcolor{blue}{(a)} shows a typical STM image of the atomically smooth Bi$_2$Te$_3$ surface. 
The corresponding height profile across the surface is plotted in Fig.  \ref{fig:STM_topography}\textcolor{blue}{(b)}. The step height of adjacent terraces is 10.3 $\text{ \AA}$. This value is consistent with one quintuple layer\cite{Zhang2009} (QL) of Bi$_2$Te$_3$ since its stacking sequence along the $\left[001\right]$ direction is Te-Bi-Te-Bi-Te, forming a QL height of 1 nm. Fig. \ref{fig:STM_topography}\textcolor{blue}{(c)} shows an atomic resolution image of our Bi$_2$Te$_3$ film surface, where the surface tellurium atoms are clearly observed exhibiting a hexagonal unit cell (see right bottom insert of Fig.  \ref{fig:STM_topography}\textcolor{blue}{(c)}). The inter-atomic spacing of these tellurium atoms was determined to be $4.3\text{ \AA}$, corresponding perfectly to the tellurium-terminated surface in  Bi$_2$Te$_3$. No surface adatoms were observed, which is an indication of a clean surface of our high quality thin films.
\begin{figure*}[!t]
 {\includegraphics[width=1\textwidth]{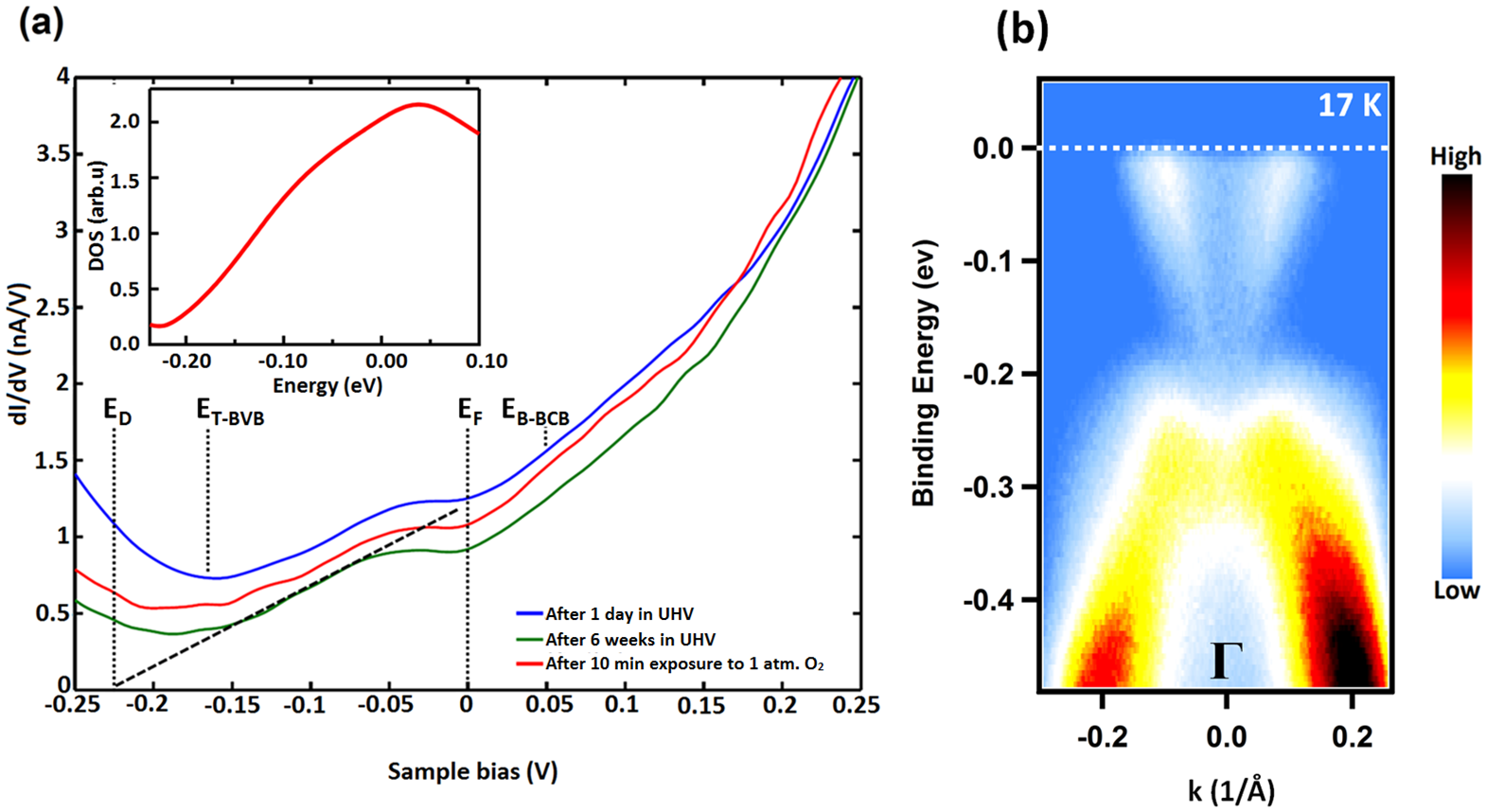}}         
  \caption{(color on-line). (a) STM spectroscopy measurements  at 77 K on a 30 nm Bi$_2$Te$_3$ film (blue curve). Effect of storing films \textit{in-situ} at $P\sim 10^{-12}$ mbar for weeks (red curve) and exposing them to pure oxygen (green curve). The inset shows calculated density of states as function of energy from the Dirac cone alone. (b) ARPES $I(k,E)$ image around the $\Gamma$ point of a 20 nm Bi$_2$Te$_3$ film recorded at 17 K. The white dashed line indicates the position of the Fermi level, which is in the bulk band gap without bulk band contribution. Both samples were grown on sapphire}
  \label{fig:STM_spectroscopy}
\end{figure*}

Now, we present spectroscopy data of Bi$_2$Te$_3$ films. Figure \ref{fig:STM_spectroscopy}\textcolor{blue}{(a)} shows low temperature differential conductivity curves of a 30 nm film grown on sapphire measured after one day in the STM chamber (blue line), after 6 weeks \textit{in-situ} storage (green) and later after being oxidized for 10 minutes in 1 atm., pure O$_2$ (red). The spectra were obtained by numerical differentiation of many I-V curves\cite{Lock-in} with setpoint parameters of $V_b= + 350$ mV at $I= 1$ nA. In this figure, E$_{\text{B-BCB}}$ marks the bottom of the bulk conduction band whereas E$_{\text{T-BVB}}$, the top of the bulk valence band. E$_{\text{F}}$ and E$_{\text{D}}$ correspond to the Fermi level and Dirac point, respectively. 
E$_{\text{B-BCB}}$ and E$_{\text{T-BVB}}$ are determined by the point where additional states appear with respect to the surface states, and by taking a literature value of $\sim 210$ meV for the band gap of Bi$_2$Te$_3$ \cite{Zhang2004,Chen2009,Zahid2010}. From these spectra, we find  linearly increasing DOS, indicative of Dirac surface states residing in the bulk energy gap between the bulk conduction band (BCB) and bulk valence band (BVB). Since the Dirac point is located in the BVB, by extrapolating this linearly increasing DOS, the Dirac point is approximately at  0.23 eV below E$_{\text{F}}$. The linear part of the spectra exhibits a plateau appearing at around $- 50$ mV. This latter feature is often attributed to the hexagonal warping of the surface band in Bi$_2$Te$_3$ samples and has been seen before in STS spectra of Bi$_2$Te$_3$ thin films and single crystals \cite{TZhang2009,Alpichshev2010,Chen2012}. A complete description of a cubic warping term in the Hamiltonian for the Dirac cone is given in Ref. \cite{Adroguer2012}, the main parameter being $\lambda$, which describes the strength of the hexagonal warping . When fitting the DOS of the warped Dirac cone to the kink feature in the data, we find that the warping alone is not strong enough (for reasonable parameter values) to explain the kink. Including a $k-$dependent tunnelling probability, implying that the states with larger parallel momenta contribute less to the tunnelling current\cite{Explanation01}, does not significantly improve the fit. However, from ARPES measurements along the $\overline{\Gamma}-\overline{M}$  direction, it is known that the Dirac cone bends outwards, as visible also in ab-initio calculations \cite{Basak2011}. Tight binding models exist that incorporate this effect \cite{Zhang2009}, and it can also be included in the Dirac Hamiltonian by including a small negative mass term. For reasonable parameter values (see details in the supplementary information) that fit the ARPES dispersion curves well, a DOS was calculated as function of energy, shown in the inset of Fig. \ref{fig:STM_spectroscopy}\textcolor{blue}{(a)}. These are the DOS from the Dirac cone alone.  It is clear that the plateau feature can indeed be qualitatively understood on the basis of the outward bending of the Dirac cone along the $\overline{\Gamma}-\overline{M}$ direction, and that it is not related to the bulk conduction band.
These STS data are in agreement with previously published STS data on Bi$_2$Te$_3$\cite{Alpichshev2010}, and indicate that there are no bulk states present at Fermi level.

To further confirm the presence of topological surface states without a contribution from the bulk bands at E$_{F}$ and also verify larger scale homogeneity of the surface states, we also performed ARPES measurements on Bi$_2$Te$_3$ films.  After growth and XPS characterization (Fig. \textcolor{blue}{S1} in the supplementary information), samples were stored in the UHV suitcase for three days and transferred to an ARPES setup. The pressure in the UHV suitcase remained lower than $2\times10^{-10}$ mbar throughout this procedure. Fig. \ref{fig:STM_spectroscopy}\textcolor{blue}{(b)} shows a representative ARPES spectrum in the vicinity of the Fermi level of a 20 nm Bi$_2$Te$_3$ film grown on sapphire around the $\Gamma$ point. The spectra were taken at a temperature of $ \sim$17  K using a helium photon source of energy 21.2 eV. The characteristic `V'-shaped surface state is clearly observed, confirming the presence of the topological Dirac cone surface state in our Bi$_2$Te$_3$ films. The white dashed line shows the position of the Fermi level, which lies well within the bulk band gap with no observable contribution of the bulk conduction band at the Fermi energy. The Fermi velocity is $\hbar v_F = 2.31\pm0.20$ eV\AA \,which is consistent with previous ARPES studies  \cite{Harrison2013,Hoefer2014} and theoretical calculations \cite{Zhang2009}. The Dirac point was measured to be at 0.24 eV below E$_{\text{F}}$, which is in good agreement with the STS  measurements for the 30 nm thick Bi$_2$Te$_3$ film shown in Fig. \ref{fig:STM_spectroscopy}\textcolor{blue}{(a)} and  previously published ARPES data \cite{Miao2012}. The fact that there are no bulk bands at E$_\text{F}$ implies that electrical conduction in these films is only due to the surface states \cite{Culcer2010}, which is desirable for electronic transport experiments.

\begin{figure}[t!]
  {\includegraphics[width=0.3\textwidth]{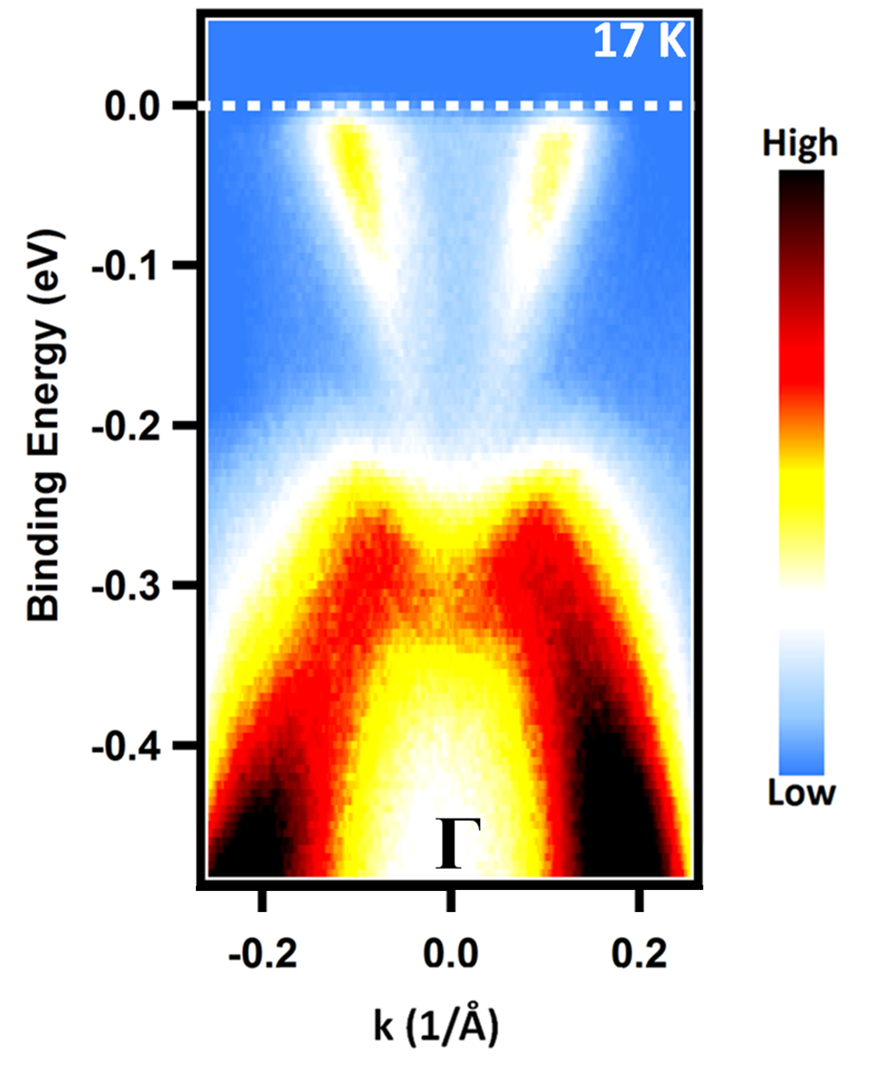}}         
  \caption{(color on-line) ARPES spectra around the $\Gamma$ point of a 15 nm Bi$_2$Te$_3$ film grown on SrTiO$_3$ [111]. As was the case for the film grown on sapphire, for this sample only the topological surface states cross the Fermi level.}
  \label{fig:ARPES_STO_111}
\end{figure}

Here, we emphasize that to stabilize the Fermi level in the band gap with no bulk bands at E$_\text{F}$, no counter-doping was necessary in our MBE-grown thin film samples. Counter-doping reduces the surface mobility and adds disorder in samples\cite{GZhang2011,Culcer2010}. To show that we have a stable and reproducible procedure for obtaining bulk insulating Bi$_2$Te$_3$ samples, we have also grown our films on STO[111] substrates. Figure \ref{fig:ARPES_STO_111} depicts an illustrative ARPES $I(k,E)$ image from a 15 nm Bi$_2$Te$_3$ film grown on STO. The data from the film grown on STO substrate agree with those from films grown on sapphire in that the Fermi level is clearly situated in the bulk band gap. This finding also confirm that despite large lattice mismatch between the film and the substrate\cite{Eibl2015}, intrinsic Bi$_2$Te$_3$ films are grown.  In particular, one can exploit these bulk insulating Bi$_2$Te$_3$ films grown on STO for fabrication of hybrid devices consisting of bottom gate-tunable topological insulator interfaced to either a superconductor or ferromagnet, by using STO [111] substrate as a back gate because it has a very high relative dielectric constant at low temperature\cite{XHe2012}

To investigate the effect of UHV storage (at room temperature) and oxidation on the band structure of our Bi$_2$Te$_3$ films, we compare STS spectra of a film measured after one day in the STM chamber, with spectra taken after 6 weeks and after being oxidized for 10 minutes at 1 atm. The oxidation was performed in the load lock of the STM system, at atmospheric pressure. From all three traces in Fig. \ref{fig:STM_spectroscopy}\textcolor{blue}{(a)} it is remarkable to observe that even after keeping films \textit{in-situ} for weeks and allowing them to interact with pure oxygen, no noticeable degradation and no corresponding shifts in the STS spectra take place. The overall key features: the linear surface band, the plateau due to `warping', and the position of the Fermi energy relative to the bulk bands, remain unchanged. This observation is further confirmed by the ARPES data shown in Fig. \ref{fig:STM_spectroscopy}\textcolor{blue}{(b)} and Fig. \ref{fig:ARPES_STO_111} since these samples were measured after storage for three days in the UHV suitcase (P$\sim 2\times10^{-10}$ mbar); and no bulk conduction bands were observed. Aging effects have been reported in ARPES data from bulk single crystals of Bi$_{2}$Te$_{3}$ within hours after cleavage under UHV conditions\cite{Chen2009,Zhou2012}. As for our films, the surface is stable as long as they are kept \textit{in-situ} in UHV conditions (see Fig. \textcolor{blue}{S6} in the supplementary information) or when exposed to pure oxygen. This conclusion is consistent with recent \textit{in-situ} four-point probe conductivity, angle resolved photoemission spectroscopy and conductive probe atomic force microscopy studies\cite{Hwang2014,Hoefer2014}.

\begin{figure}[!t]
 {\includegraphics[width=0.4\textwidth]{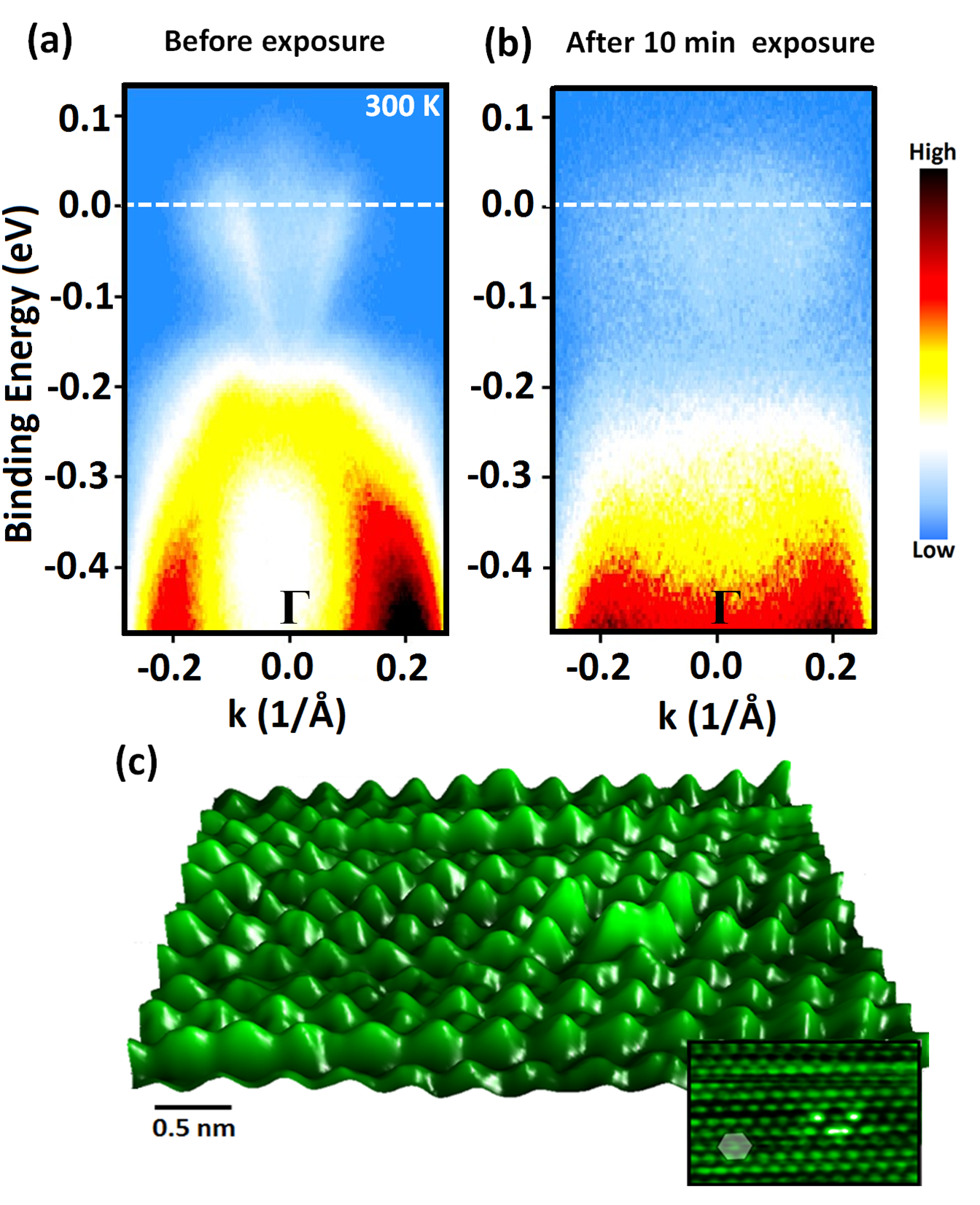}}         
  \caption{(color on-line) Effect of air exposure under ambient conditions on spectroscopy and topography of Bi$_2$Te$_3$ films grown on sapphire. (a) ARPES spectra around the $\Gamma$ point of a 20 nm film before and (b) after 10 minuntes exposure to atmospheric pressure. (c) STM topography ($6\times 3.5\text{nm}^2$) of a 30 nm film after exposure to ambient conditions. The surface tellurium atoms with a hexagonal crystal structure (right bottom insert) are still resolved together with noticeable defects as compared to before exposure (same sample as in Fig. \ref{fig:STM_topography}\textcolor{blue}{(c)}) to ambient conditions.}
  \label{fig:HexagonalWrappingEffect}
\end{figure}

Next, we study the effect of exposure to ambient conditions. This is relevant for the understanding of how \textit{ex-situ} contamination processes affect surface and bulk states of Bi$_2$Te$_3$ films for future \textit{ex-situ} fabricated hybrid topological insulators devices. For this investigation, we employ both STM/STS and ARPES since they are complementary tools for probing reliably the band structure of materials. Figure \ref{fig:HexagonalWrappingEffect}\textcolor{blue}{(a)} depicts an ARPES $I(k,E)$ image of the 20 nm Bi$_2$Te$_3$ film grown on sapphire. This time, the sample was measured at room temperature before exposure to ambient conditions. This $I(k,E)$ image was recorded after the sample had been in the ARPES chamber for 5 days, after measurement of the low-T data shown in Fig. \ref{fig:STM_spectroscopy}\textcolor{blue}{(b)} (which themselves were after three days in UHV suitcase and transfer to ARPES setup). The Fermi level is still lying below the bottom of the bulk conduction band (B-BCB). From this room temperature ARPES spectra, states up to $\sim$50 meV  above the Fermi level are resolved due to thermal population. We took this 20 nm film measured \textit{in-situ}, both at room and low temperature, out of the ARPES system and exposed it to air for 10 minutes. After this exposure, we loaded the sample again (no further treatments such as annealing of the sample) and reacquired ARPES data in order to assess the influence of air exposure on the band structure of our Bi$_2$Te$_3$ films. We can clearly see from Fig. \ref{fig:HexagonalWrappingEffect}\textcolor{blue}{(b)} that bulk valence bands have moved downward after exposure and compared to the pristine sample, evident changes in the linearly dispersing Dirac surface states are observed. There are no surface states resolved after exposure to air, implying that the film degradation is deeper than the probing depth of the ARPES experiment ($\sim$1 nm). Thus, the conductance of the films will be influenced by this downward shift of bulk bands, and the corresponding upward shift of E$_\text{F}$. Recently, upward shift of E$_\text{F}$ in Bi$_2$Te$_3$ films exposed to air has also been reported in ARPES study\cite{Hoefer2014}.

We now discuss the effect of exposure to air in STM/STS data. We took the previously \textit{in-situ} measured 30 nm Bi$_2$Te$_3$  film grown on sapphire out of the STM system and exposed it to air for 10 minutes. After this procedure, we acquired several topography and spectroscopy maps at different position on the exposed surface. We used the same setpoint parameters as for the spectra shown in Fig.    \ref{fig:STM_spectroscopy}\textcolor{blue}{(a)}. We observe once more clear changes in the general shape of the spectra (Fig. \textcolor{blue}{S5} in the supplementary information): most notably, the region previously associated with linearly dispersing surface bands is absent in the exposed sample. This observation is consistent with our ARPES data of the exposed 20 nm film (see Fig.  \ref{fig:HexagonalWrappingEffect}\textcolor{blue}{(b)}) since there also, a clear degradation of the Dirac surface states was observed. 

Based on a comparison of measured STM topographies before and after exposure, we find indications that extrinsic defects due to contamination are responsible for the observed degradation of the surface states in the spectroscopic measurements. Figure  \ref{fig:HexagonalWrappingEffect}\textcolor{blue}{(c)} shows atomically-resolved STM topography image of a 30 nm Bi$_2$Te$_3$ film after exposure to ambient conditions, in which additional defects can be readily identified. On top of the observed defects and the observed disappearance of surface states, also shifts in the position of the bulk bands due to band bending are observed in the STS spectra (see Fig. \textcolor{blue}{S5} in the supplementary information). Whether bands shift upward or downward might sensitively depend on the exact exposure conditions as well as the position where spectra are taken (e.g. close to a defect). Nonetheless, from the combined ARPES and STS spectroscopic data it is clear that exposure to ambient pressures changes notably the surface states of our Bi$_2$Te$_3$ thin films. These results underline the need of protecting films before taking them \textit{ex-situ} for further studies, since extrinsic defects may be responsible for the change of sign of the carriers in the surface conductivity of TIs \cite{Taskin2011}; for the deterioration of the topological surface state properties \cite{Kong2011} and the reduction of the spin polarization \cite{Wang2012}.

The XPS surface analysis of the exposed sample reveals a splitting of the Te $3d_{3/2}$ and Te $3d_{5/2}$ peaks consistent with the formation of tellurium oxide\cite{Explanation03} (TeO$_2 $), see Fig. \textcolor{blue}{S2(b)} and \textcolor{blue}{S2(c)}  in the supplementary information. Since it was already shown that only pure oxygen doesn't affect topological surface states (see above and Ref.\cite{Hoefer2014}) we believe that  other molecules from atmosphere, most likely water vapour, are the origin of the observed contamination. This conclusion emphasizes that for further \textit{ex-situ} electronic transport investigations and quantum device fabrication, films should first be capped \textit{in-situ} to avoid any possible contamination as has already been indicated for Bi$_2$Se$_3$ thin films \cite{Dai2014}. For Bi$_2$Te$_3$ films, though, a systematic study is needed in order to determine a suitable capping layer that can preserve the surface states and avoid unintentional doping induced by ambient conditions. An insulating layer would be appropriate since it will allow the fabrication of nearly damage-free top gate structures, so that the carriers in the top surface can be efficiently tuned \cite{Yang2014}

In summary, we have systematically investigated the effect of \textit{in-situ} exposure to pure oxygen at atmospheric pressure and of \textit{ex-situ} contamination on high quality intrinsic Bi$_2$Te$_3$ films grown sapphire and STO. The XPS studies shows narrow and symmetric Te and Bi peaks highlighting the absence of Te or Bi excess in our films. STM confirms that the surface has atomically flat terraces and is free from intrinsic defects. The analysis of STS spectra underlines that the surface states of Bi$_2$Te$_3$ films are stable against oxidation and do not degrade as long as they are stored \textit{in-situ} in ultra high vaccuum or in pure oxygen. ARPES data recorded from films grown in the same manner further confirms the presence of topological surface states with no bulk bands at the Fermi level. This study also provides a platform to investigate bottom gate tunable hybrid transport\cite{Kurter2014,Morpurgo2011} at the surface of intrinsic Bi$_2$Te$_3$ films grown on STO [111] substrates. The combined \textit{in-} and \textit{ex-situ} spectroscopy study reported here emphasizes the importance of capping these films before exposure to subsequent device fabrication steps. A possible route is to \textit{in-situ} cap the Bi$_2$Te$_3$ and other TI films with an insulating AlO$_\text{x}$ layer.
\newline

We thank Dominic Post and Frank Roesthuis for their technical support in upgrading the growth system. This work is financially supported by the Dutch Foundation for Fundamental Research on Matter (FOM), the Netherlands Organization for Scientific Research (NWO) and by the European Research Council (ERC).


\begin{thebibliography}{1}
\bibitem{Fu2007} L. Fu, C. L. Kane, and E. J. Mele, Phys. Rev. Lett. \textbf{98}, 106803 (2007).
 \bibitem{Zhang2009}H. Zhang, C. -X. Liu, X. -L Qi, X. Dai, Z. Fang, S. -C. Zhang, Nature Phys. \textbf{5}, 438 (2009).
 \bibitem{Moore2010} J. E. Moore, Nature \textbf{464}, 194 (2010).
\bibitem{He2012}
L. He, F. Xiu, X. Yu, M. Teague, W. Jiang, Y. Fan, X. Kou, M. Lang, Y. Wang, G. Huang, N. -C. Yeh, and K. L. Wang, Nano Lett. \textbf{12}, 1486 (2012).
\bibitem{Taskin2012} A. A. Taskin, S. Sasaki, K. Segawa, and Y. Ando, Phys. Rev. Lett. \textbf{109}, 066803 (2012).
\bibitem{He2013}
L. He, X. Kou, and K. L. Wang, Phys. Status Solidi RRL \textbf{7}, 50 (2013).
\bibitem{Li2010}
Y. Li, G. Wang, X. Zhu, M. Liu, C. Ye, X. Chen, Y. Wang, K. He, L. Wang, X. Ma, H. Zhang, X. Dai, Z. Fang, X. Xie, Y. Liu, X. Qi, J. Jia, S. Zhang, and Q. Xue, Adv. Funct. Mater.  \textbf{22}, 4002 (2010).
\bibitem{Wang2011}
G. Wang, X. Zhu, Y. Sun, Y. Li, T. Zhang, J. Wen, X. Chen, K. He , L. Wang, X. Ma, J. Jia, S. B. Zhang, and Q. Xue, Adv. Funct. Mater.  \textbf{23}, 2929 (2011).
\bibitem{Hoefer2014}
K. Hoefer, C. Becker, D. Rata, J. Swanson, P. Thalmeier, and L. H. Tjeng, Proc. Natl. Acad. Sci. USA \textbf{111}, 14979 (2013).
\bibitem{XHe2012}
X. He, T. Guan, X. Wang, B. Feng, P. Cheng, L. Chen,
Y. Li, and K. Wu, App. Phys. Lett. \textbf{101}, 123111 (2012).
\bibitem{Benia2011}
H. M. Benia, C. Lin, K. Kern, and C. R. Ast,  Phys. Rev. Lett. \textbf{107}, 177602 (2011).
\bibitem{CChen2012}
C. Chen, S. He, H. Weng, W. Zhang, L. Zhao, H. Liu, X. Jia, D. Mou,
S. Liu, J. He, Y. Peng, Y. Feng, Z. Xie, G. Liu, X. Dong, J. Zhang,
X. Wang, Q. Peng, Z. Wang, S. Zhang, F. Yang, C. Chen,
Z. Xu, X. Dai, Z. Fang, and X. J. Zhou, Proc. Natl. Acad. Sci. USA \textbf{109}, 3694 (2012).
\bibitem{Eibl2015}
O. Eibl, K. Nielsch, N. Peranio and F. V\"{o}lklein, \textit{Thermoelectric Bi$_2$Te$_3$ Nanomaterials}, Wiley-VCH Verlag GmBh \& Co. KGaA, Weinheim, Germany (2015).
\bibitem{Chang2008}
J. Chang, Y.-S. Park, and S.-K. Kim, Appl. Phys. Lett. \textbf{92}, 152910 (2008).
\bibitem{GZhang2011}
G. Zhang , H. Qin , J. Chen , X. He , L. Lu , Y. Li , and K. Wu,
Adv. Funct. Mater.  \textbf{21}, 2351 (2011).
\bibitem{Bansal2011} N. Bansal, Y. S. Kim, E. Edrey, M. Brahlek, Y. Horibe, K. Iida, M. Tanimura, G. -H. Li, T. Feng, H. -D. Lee, T. Gustafsson, E. Andrei, and S. Oh, Thin Solid Films \textbf{520}, 224 (2011).
\bibitem{Harrison2013} S. E. Harrison, S. Li, Y. Huo, B. Zhou, Y. L. Chen, and J. S. Harris, Appl. Phys. Lett. \textbf{102}, 171906 (2013).
\bibitem{Bando2000}
H. Bando, K. Koizumi, Y. Oikawa, K. Daikohara, V. A. Kulbachinskii and H. Ozaki, J. Phys.: Condens. Matter \textbf{12}, 5607 (2000). 
\bibitem{Kong2011}
D. Kong, J. J. Cha, K. Lai, H. Peng, J. G. Analytis, S. Meister,
Y. Chen, H. Zhang, I. R. Fisher, Z. Shen, and Y. Cui, ACS Nano \textbf{5}, 4698 (2011).
\bibitem{Guo2013}
J. Guo, F. Qiu, Y. Zhang, H. Deng, G. Hu, X. Li, G. Yu and N. Dai, Chin. Phys. Lett. \textbf{30}, 106801 (2013).
\bibitem{Yashina2013}
L. V. Yashina, J. S\'{a}nchez-Barriga, M. R. Scholz, A. A. Volykhov, A. P. Sirotina, V. S. Neudachina, M. E. Tamm, A. Varykhalov,D. Marchenko, G. Springholz, G. Bauer, A. Knop-Gericke and O. Rader, ACS Nano \textbf{7}, 5181 (2013).
\bibitem{Roy2013}
A. Roy, S. Guchhait, S. Sonde, R.Dey, T. Pramanik,
A. Rai, H. C. P. Movva, L. Colombo, and S. K. Banerjee, Appl. Phys. Lett. \textbf{102}, 163118 (2013).
\bibitem{Zhang2004} H.T. Zhang, X.G. Luo, C.H. Wang, Y.M. Xiong, S.Y. Li, X.H. Chen, J. Crys. Grow. \textbf{265}, 558   (2004).
\bibitem{Park2012}
J. Park, Y. -A Soh, G. Aeppli, S. R. Bland, X. -G. Zhu, X. Chen, Q. -K. Xue, and F. Grey, Appl. Phys. Lett. \textbf{101}, 221910  (2012).
\bibitem{Koma1992_1}
A. Koma, Surf. Sci. \textbf{267}, 29 (1992).
\bibitem{Koma1992_2}
A. Koma, Thin Solid Films \textbf{216}, 72 (1992).
\bibitem{Firpo2004} G. Firpo, and A. Pozzo, Rev. Sci. Instrum. \textbf{75}, 4828 (2004).
\bibitem{Urazhdin2004} S. Urazhdin, D. Bilc, S. D. Mahanti, and S. H. Tessmer, Phys. Rev. B \textbf{69}, 085313 (2004). 
\bibitem{Alpichshev2010} Z. Alpichshev, J. G. Analytis, J. -H. Chu, I. R. Fisher, Y. L. Chen, Z. X. Shen, A. Fang, and A. Kapitulnik, Phys. Rev. Lett. \textbf{104}, 016401 (2010).
\bibitem{Chen2012} M. Chen, J. Peng, H. Zhang, L. Wang, K. He, X. Ma, and Q. Xue, Appl. Phys. Lett. \textbf{101}, 081603 (2012).
\bibitem{Lock-in} To get differential conductivity spectra shown in this work, the sum of more than 3600 I-V curves was numerically differentiated. The I-V spectra were also recorded by means of the standard lock-in technique with a lock-in time constant of $10$ ms. The lock-in extracted dI/dV spectra were consistent to the ones obtained with numerical differentiation.
\bibitem{Chen2009}Y. L. Chen, J. G. Analytis, J. -H. Chu, Z. K. Liu, S. -K. Mo, X. L. Qi, H. J. Zhang, D. H. Lu, X. Dai, Z. Fang, S. C. Zhang, I. R. Fisher, Z. Hussain and Z. -X. Shen, Science  \textbf{325}, 178  (2009).
\bibitem{Zahid2010}F. Zahid and R. Lake,  Appl. Phys. Lett. \textbf{97}, 212102 (2010).
\bibitem{TZhang2009} T. Zhang, P. Cheng, X. Chen, J. F. Jia, X. Ma, K. He, L. Wang, H. Zhang, X. Dai, Z. Fang, X. Xie, and Q. -K. Xue, Phys. Rev. Lett. \textbf{103}, 266803 (2009).

\bibitem{Adroguer2012}P. Adroguer, D. Carpentier, J. Cayssol and E. Orignac, New J. Phys. \textbf{14}, 103027 (2012).
\bibitem{Explanation01}
To perform this analysis, we added a probability factor of the form (see Ref. \cite{Houselt2008,Zandvliet2009} for details): $P\propto e^{2z}\sqrt{C+k^2_{\parallel}}$, where $z$ is the tip-sample separation in the tunnelling experiment, and $C$ the height of the tunnelling barrier in units of $k^2$. Nonetheless, this improved fitting yielded insufficient result since the modelled kink due to warping was not strong enough to capture the measured kink feature in conductance spectra.

\bibitem{Houselt2008} A. van Houselt, T. Gnielka, M. Fischer, J. M. J. Aan de Brugh, N. Oncel, D. Kockmann, R. Heid, K. -P. Bohnen, B. Poelsema and H. J. W. Zandvliet, Surf. Sci. \textbf{602}, 1731 (2008).
\bibitem{Zandvliet2009}H. J. W. Zandvliet and A. van Houselt, Annu. Rev. Anal. Chem.  \textbf{2}, 37 (2009).
\bibitem{Basak2011} S. Basak, H. Lin, L. A. Wray, S.-Y. Xu, L. Fu, M. Z. Hasan, and A. Bansil, Phys. Rev. B \textbf{84}, 121401(R) (2011).
\bibitem{Miao2012}L. Miao, Z. F. Wang, W. Ming, M. -Y. Yao, M. Wang, F. Yang, Y. R. Songa, F. Zhu, A. V. Fedorovc, Z. Sund, C. L. Gao, C. Liu, Qi-Kun Xuee, C. -X. Liuf, F. Liub, D. Qian,
and J. -F. Jia,  Proc. Natl. Acad. Sci. USA \textbf{110}, 2758 (2012).
\bibitem{Culcer2010}
D. Culcer, E. H. Hwang, T. D. Stanescu, S. Das Sarma,. Phys. Rev. B \textbf{82}, 155457 (2010).
\bibitem{Zhou2012}B. Zhou, Z. K. Liu, J. G. Analytis, K. Igarashi, S. K. Mo, D. H. Lu, R. G. Moore, I. R. Fisher, T. Sasagawa, Z. X. Shen, Z. Hussain and Y. L. Chen, Semicond. Sci. Technol. \textbf{27}, 124002 (2012).
\bibitem{Hwang2014}J. H. Hwang, J. Park, S. Kwon, J. S. Kim and J. Y. Park, Surf. Sci. \textbf{630}, 153 (2014).
\bibitem{Taskin2011} A. A. Taskin, Z. Ren, S. Sasaki, K. Segawa, and Y. Ando, Phys. Rev. Lett. \textbf{107}, 016801 (2011).
\bibitem{Wang2012} X. Wang, G. Bian, T. Miller, and T. -C. Chiang, Phys. Rev. Lett. \textbf{108}, 096404 (2012).
\bibitem{Explanation03}
To find the stoichiometry of the TeO$_\text{x}$ layer, we fitted the oxidized XPS spectra using the same fitting procedure as for the XPS spectra in Fig. \ref{fig:XPS}. The TeO$_\text{x}$ peak was found at $ 576.01\pm0.05$ eV, which is in agreement with the peak position of TeO$_2$ given in the Handbook of x-ray photoelectron spectroscopy\cite{Moulder1995}.
\bibitem{Moulder1995}
J. F. Moulder, W. F. Stickle, P. E. Sobol, and K. D. Bomben, \textit{Handbook of X-ray Photoelectron Spectroscopy}, Physical Electronics, Inc., Minnesota, USA (1995).
\bibitem{Dai2014}J. Dai, W. Wang, M. Brahlek, N.  Koirala, M. Salehi, S. Oh and W. Wu, Nano Research, 1-7 (2014), doi:10.1007/s12274-014-0607-8.

\bibitem{Yang2014} F. Yang, A. A. Taskin, S. Sasaki, Ko. Segawa, Y. Ohno, K. Matsumoto, and Y. Ando, Appl. Phys. Lett. \textbf{104}, 161614 (2014).
\bibitem{Kurter2014}
C. Kurter, A. D. K. Finck, P. Ghaemi, Y. S. Hor, and D. J. Van Harlingen, Phys. Rev. B \textbf{90}, 014501 (2014). 
\bibitem{Morpurgo2011}
B. Sac\'ep\'e, J.  B. Oostinga, J. Li, A. Ubaldini , N. J. G. Couto, E. Giannini and A. F. Morpurgo, Nat. Commun. \textbf{2}, 575 (2011).
\end{thebibliography}
\end{document}


\title{In-situ spectroscopy of intrinsic Bi$_2$Te$_3$ topological insulator thin films and impact of extrinsic defects: Supplementary Information}
\author{P. Ngabonziza}
\affiliation{Faculty of Science and Technology and MESA+ Institute for Nanotechnology, University of Twente, 7500 AE Enschede, The Netherlands}

 \author{R. Heimbuch}
 \affiliation{Faculty of Science and Technology and MESA+ Institute for Nanotechnology, University of Twente, 7500 AE Enschede, The Netherlands}
 \author{ N. de Jong}
\affiliation{Van der Waals-Zeeman Institute, University of Amsterdam, Science Park 904, 1098 XH, Amsterdam, Netherlands}
 \author{R. A. Klaassen}
\affiliation{Faculty of Science and Technology and MESA+ Institute for Nanotechnology, University of Twente, 7500 AE Enschede, The Netherlands}
\author{M. P. Stehno}
\affiliation{Faculty of Science and Technology and MESA+ Institute for Nanotechnology, University of Twente, 7500 AE Enschede, The Netherlands}
\author{M. Snelder}
\affiliation{Faculty of Science and Technology and MESA+ Institute for Nanotechnology, University of Twente, 7500 AE Enschede, The Netherlands}
\author{A. Solmaz}
\affiliation{Faculty of Science and Technology and MESA+ Institute for Nanotechnology, University of Twente, 7500 AE Enschede, The Netherlands}
\author{S. V. Ramankutty}
\affiliation{Van der Waals-Zeeman Institute, University of Amsterdam, Science Park 904, 1098 XH, Amsterdam, Netherlands}
\author{E. Frantzeskakis}
\affiliation{Van der Waals-Zeeman Institute, University of Amsterdam, Science Park 904, 1098 XH, Amsterdam, Netherlands}
\author{ E. van Heumen}
\affiliation{Van der Waals-Zeeman Institute, University of Amsterdam, Science Park 904, 1098 XH, Amsterdam, Netherlands}
\author{G. Koster}
\affiliation{Faculty of Science and Technology and MESA+ Institute for Nanotechnology, University of Twente, 7500 AE Enschede, The Netherlands}
\author{M. S. Golden}
\affiliation{Van der Waals-Zeeman Institute, University of Amsterdam, Science Park 904, 1098 XH, Amsterdam, Netherlands}
\author{H. J. W. Zandvliet}
\affiliation{Faculty of Science and Technology and MESA+ Institute for Nanotechnology, University of Twente, 7500 AE Enschede, The Netherlands}
\author{A. Brinkman}
\affiliation{Faculty of Science and Technology and MESA+ Institute for Nanotechnology, University of Twente, 7500 AE Enschede, The Netherlands}
\date{\today}
\pacs{79.60.Dp, 73.20.-r, 68.37.Ef, 79.60.-i}

\maketitle
\section{XPS measurements}
X-ray photoemission spectroscopy (XPS) spectra were taken \textit{in-situ} using an Omicron Nanotechnology system equipped with a monochromatic aluminum source (K$\alpha$  x-ray source $\text{XM}1000$) with a photon energy of $1486.7$ eV.  The background pressure was $\sim 5\times10^{-11}$ mbar. The kinetic energy of electrons emitted from the sample were analyzed using a 7 channel EA 125 electron analyzer operated in CAE mode. XPS measurements were performed a few hours after film growth, with the films having remained in UHV at all times. 
\begin{figure*}[!h]
{\includegraphics[width=1\textwidth]{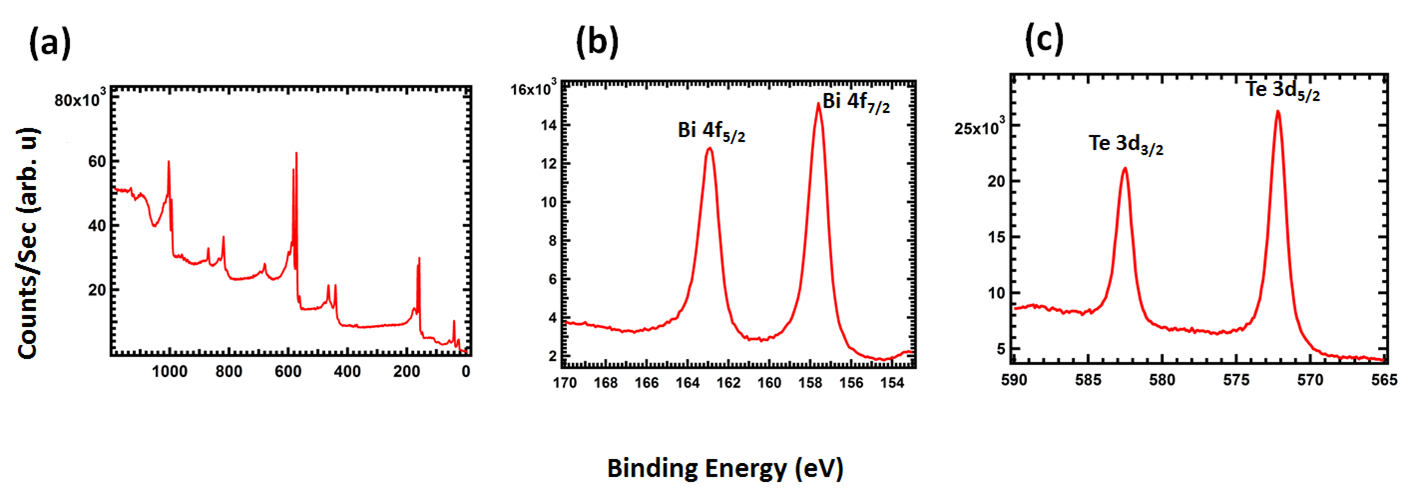}}         
\textbf{Figure S1:} (color on-line) \textit{In-situ} XPS measurement on a 20 nm Bi$_2$Te$_3$ film grown on sapphire that was subsequently measured in ARPES experiment (see Fig. \textcolor{blue}{3(b)}). (a) Survey scan taken a few hours after the film growth. Only Bi and Te peaks are resolved. (b) and (c) High-resolution scans around the Bi 4$f$ and Te 3$d$ main peaks, respectively.
\end{figure*}
\begin{figure*}[!h]
{\includegraphics[width=1\textwidth]{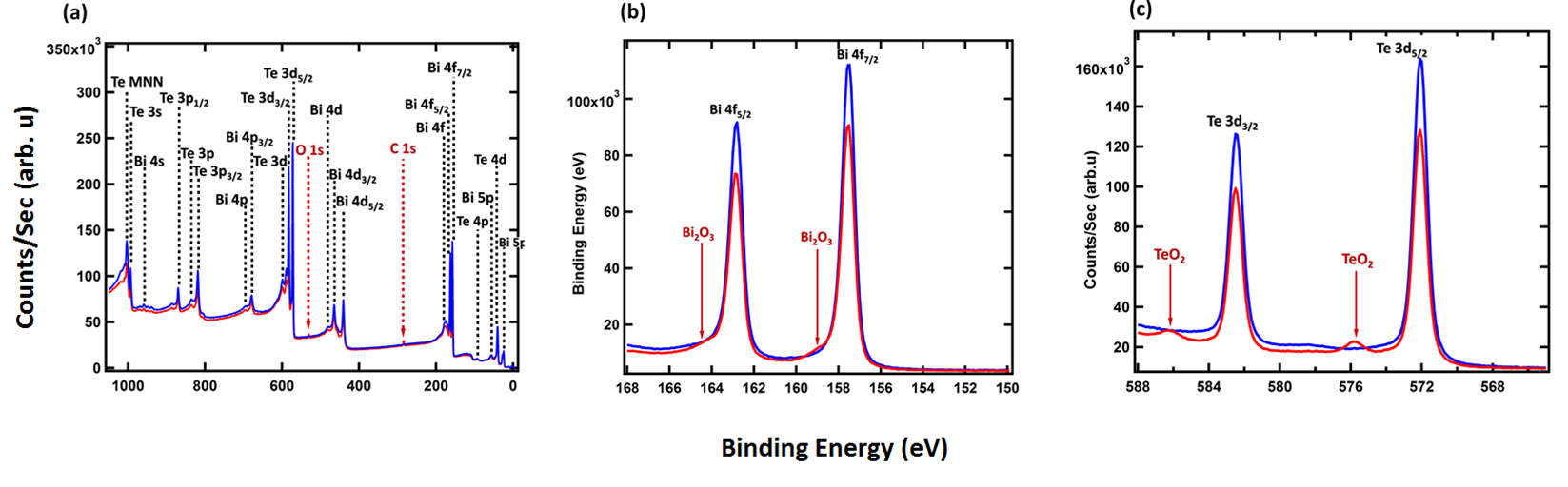}}         
\textbf{Figure S2:} (color on-line) Effect of \textit{ex-situ} exposure on a 30 nm Bi$_2$Te$_3$ film grown on sapphire. (a) An XPS survey scan when the sample is measured \textit{in-situ} (blue) and after being exposed to ambient conditions (red). After \textit{ex-situ} exposure, oxygen and carbon peaks appear. High resolution Bi 4$f$ (b) and Te 3$d$ (c) XPS spectra before and after short exposure. After exposure, there is a clear formation of a TeO$_2$ layer, whereas Bi$_2$O$_3$ formation is barely noticeable. This is consistent with the fact that the surface in an ideal crystal is terminated by tellurium atoms\cite{Zhang2009}. The bismuth oxidation would be expected to become significant, after the tellurium outer layer has been completely oxidized.
\end{figure*}
\newpage
\section{XRD measurements }
To further characterize our films, additional \textit{ex-situ} x-ray diffraction (XRD) measurements were done, which indicated the high crystalline quality of our Bi$_2$Te$_3$ films, grown using the two-step procedure described in the main body of the paper. We show below XRD data of a 30 nm film grown on Al$_2$O$_3$ [0001] and a 15 nm film grown on STO [111]. The film on sapphire was measured at low temperature using STM/STS, whereas the film on STO was measured using ARPES (See main text).
\begin{figure}[!h]
{\includegraphics[width=0.9\textwidth]{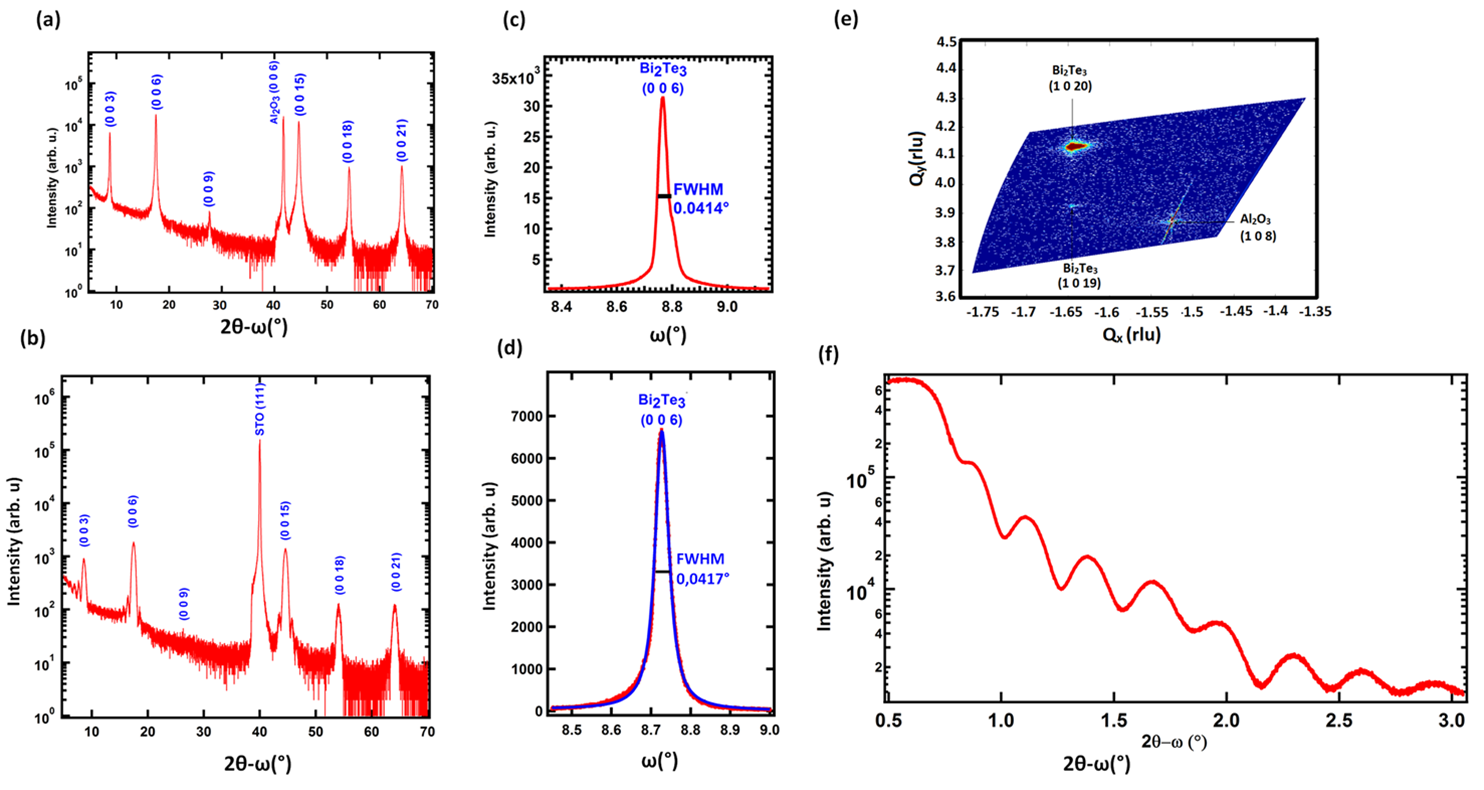}}

\textbf{Figure S3: } (color on-line) \textit{Ex-situ} XRD measurements. (a) and (b): $2\theta-\omega$ scans of Bi$_2$Te$_3$ films grown on sapphire and STO. Only the substrate 
peaks and the (0 0 3) family of Bi$_2$Te$_3$ diffraction peaks are resolved. (c) and (d): Rocking curves of the Bi$_2$Te$_3$ main peak at (0 0 6). (e) A 2D reciprocal space map of a 30 nm film grown on sapphire. Positions of the Al$_2$O$_3$ (1 0 8) and Bi$_2$Te$_3$ (1 0 20) and (1 0 19) peaks are shown. (f) X-ray reflectivity data of a 15 nm film grown on STO showing distinct Kiessig fringes over more than five orders of magnitude in intensity. Despite the significant lattice mismatch between the film and substrates\cite{He2013,Eibl2015}, high structural-quality films are grown.
  \label{fig:XPS}
\end{figure}
\newpage
\section{Model calculation of density of states and dispersion relations}
To model Bi$_{2}$Te$_{3}$ dispersion relations along the $\overline{\Gamma}-\overline{M}$ and  $\overline{\Gamma}-\overline{K}$  directions, we used the k$\cdot$p Hamiltonian described in Ref. \cite{Zhang2009, Basak2011} to calculate the Green's function and subsequently the Density Of States (DOS). This model includes anisotropic higher order terms, for example describing the outward bending of the Dirac cone along the $\overline{\Gamma}-\overline{M}$ direction at the energy close to the maximum of the valence band, also present in ab-initio calculations \cite{Basak2011}. We used as few parameters as possible to describe the measured ARPES dispersions. The parameters used are (in the same notation as Ref. \cite{Basak2011}): $m^{*}_{1}$ = -0.054 $eV^{-1}\dot{A}^{-2}$, $v$ = 2.55 $eV\dot{A}$, $\alpha$ = 5.5 $eV\dot{A}^{2}$, $\lambda$ = 250 $eV\dot{A}^{3}$ and the chemical potential $\mu$ = 225 $meV$ above the Dirac point, which coincides with the data from the STS experiments. The Green's function was calculated on a grid of 1200 $\times$ 1200 $k$-points with an energy step of 3 meV.
Figure \textcolor{blue}{S4 (a)} and \textcolor{blue}{(b)} show the spectral functions obtained by summing the imaginary parts of the diagonal components of the Green's function. The DOS, shown as inset in Fig. \textcolor{blue}{3 (a)} in the main text, was obtained by integrating the spectral functions over the Brillouin zone.
\begin{figure*}[h!]
{\includegraphics[width=0.8\textwidth]{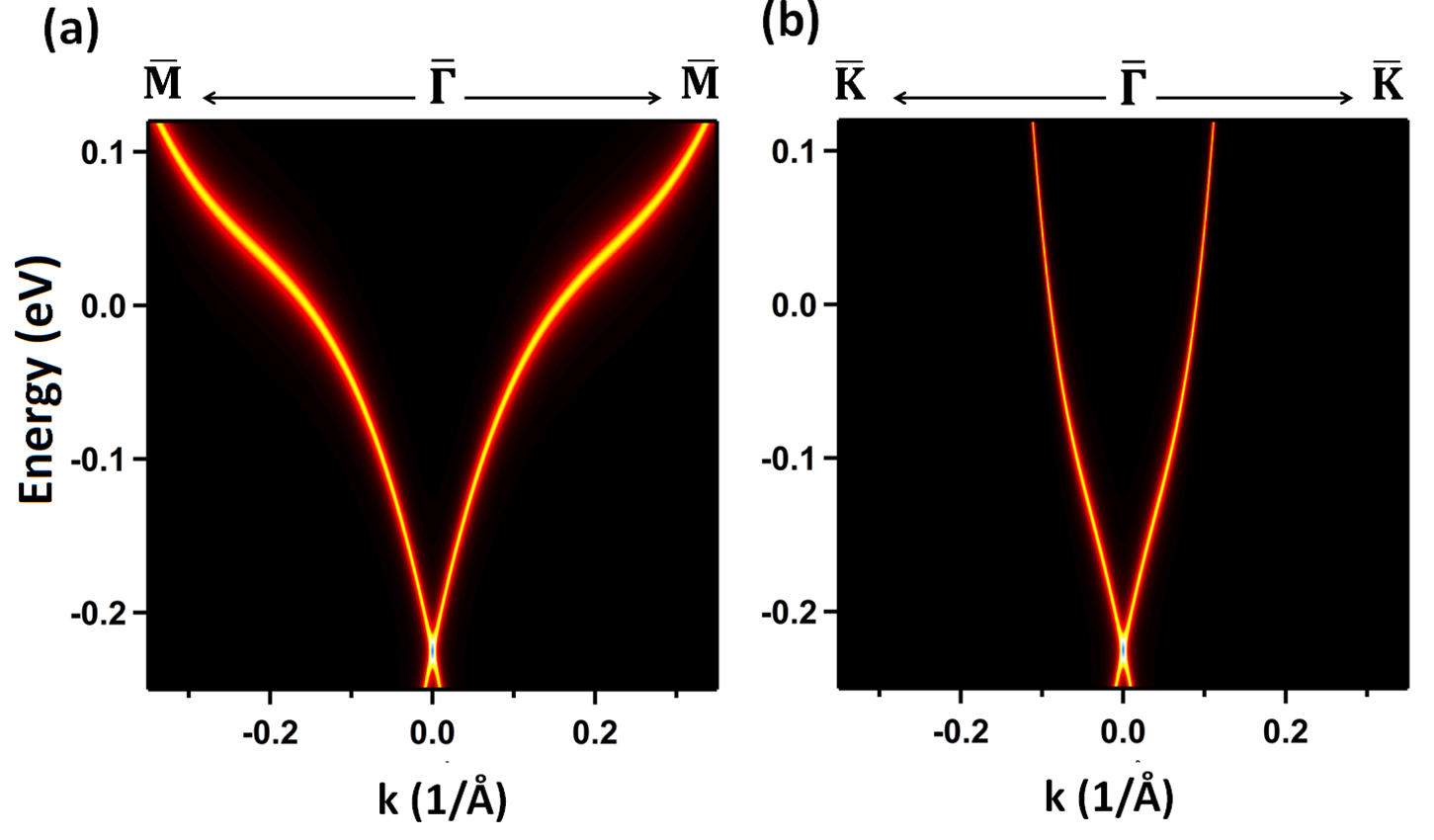}}         

\textbf{Figure S4:}\ (color on-line) Calculated spectral functions of Bi$_{2}$Te$_{3}$ along the (a) $\overline{\Gamma}-\overline{M}$ and (b) $\overline{\Gamma}-\overline{K}$ directions. Outward bending of the Dirac cone is observed only along the $\overline{\Gamma}-\overline{M}$ direction.
  \label{fig:DOS1}
\end{figure*}
\newpage
\section{STM spectroscopy before and  after exposure to ambient conditions}
To investigate the impact of exposure to ambient conditions, the 30 nm Bi$_{2}$Te$_{3}$ film grown on sapphire was exposed for 10 minutes to ambient conditions, then loaded back into the STM chamber. After this procedure, we took several topography images at different positions on the exposed surface. We used the same setpoint parameters as for the spectra discussed in the main text. The differential conductance curves before and after atmospheric exposure are plotted together below. Clear changes in the general shape of the spectra are observed, as was also the case for the ARPES spectra of the film exposed to air (Fig. \textcolor{blue}{5 (b)} in the main text and related discussions).
\begin{figure*}[h!]
{\includegraphics[width=0.6\textwidth]{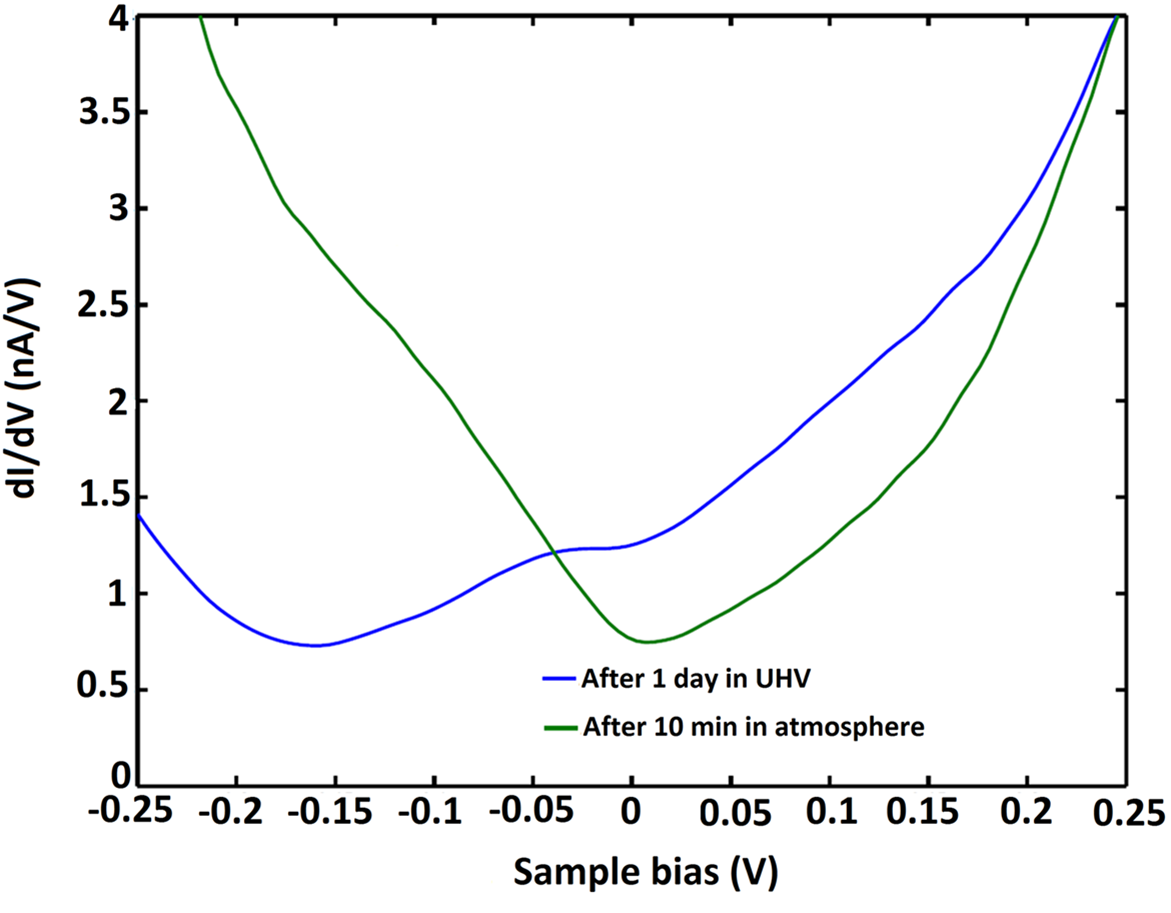}}         
 
 \textbf{Figure S5:} (color on-line) Conductance spectra of a film exposed for 10 min exposed to ambient (air) conditions. Before exposure (blue) and after exposure (green). The region previously associated with linearly dispersing surface bands is absent in the exposed sample, highlighting the deterioration of surface states after exposure to ambient conditions as observed in the ARPES data (see Fig. \textcolor{blue}{5(b)} in the main text).
  \label{fig:STS_Exposure}
\end{figure*}
\newpage
\section{ARPES measurements}
ARPES spectra in the vicinity of the Fermi level were also acquired on Bi$_{2}$Te$_{3}$ films grown on SrTiO$_{3}$ [111] substrates. The spectra were taken both at low ($\sim 17$K) and room temperature using a Scienta VUV5000 helium photon
source of 21.2 eV photon energy (He I line) and 40.8 eV (He II line) for valence band spectra; and a Scienta 2002 hemispherical electron analyzer.
 
Figure \textcolor{blue}{S6 (a)} shows an illustrative ARPES spectra of a 15 nm Bi$_{2}$Te$_{3}$ film grown on STO in two steps growth at a substrate temperatures of 190\degree C (first layer) and 250\degree C (second layer). Spectra were taken at low temperature. Only topological surface states intersecting the Fermi level are observed for this sample also. To investigate the effect of UHV exposure on the electronic structure of these films, we performed ARPES measurements over different time intervals in UHV condition, with the results shown in Fig. \textcolor{blue}{S6 (a)}-\textcolor{blue}{(d)}.
\begin{figure*}[!h]
{\includegraphics[width=1\textwidth]{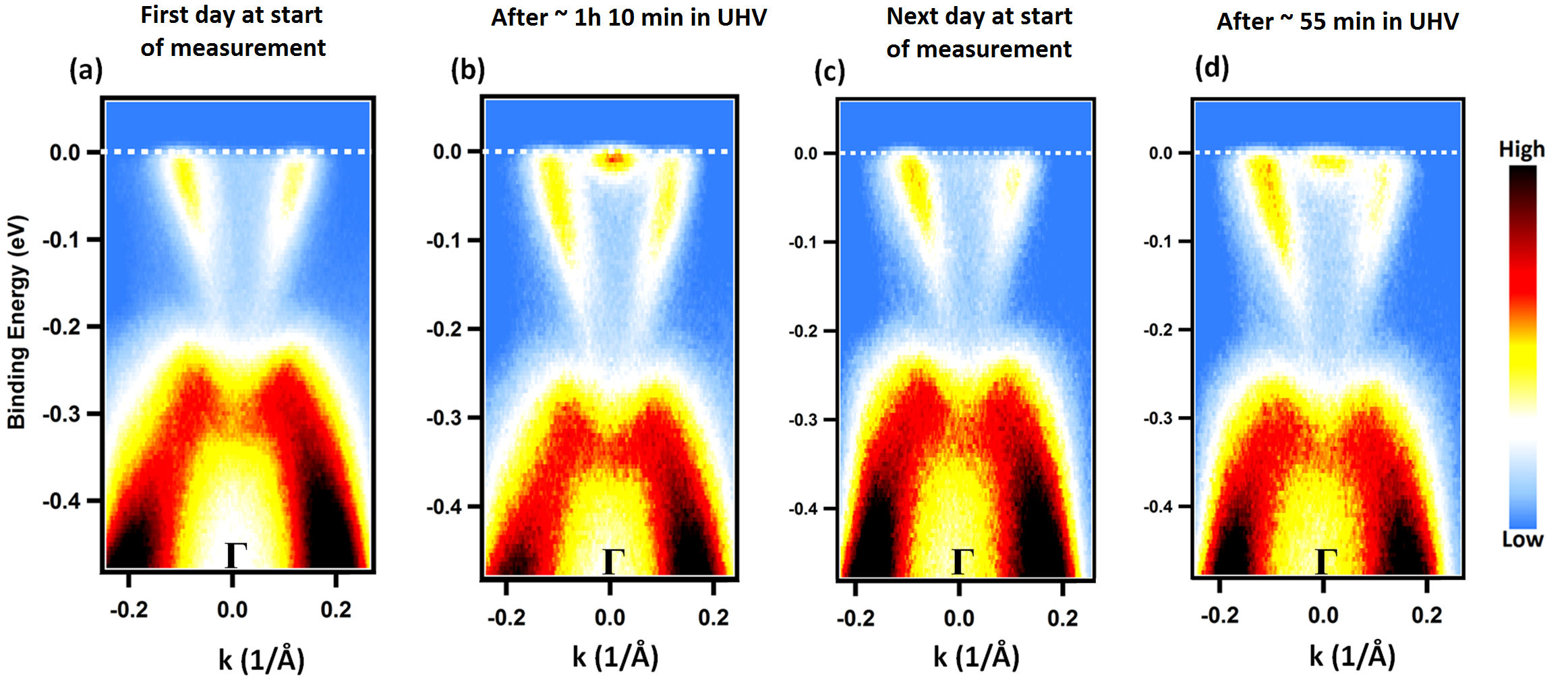}}         
 \textbf{Figure S6:}
ARPES spectra of a 15 nm thick  Bi$_{2}$Te$_{3}$ film grown on SrTiO$_{3}$ [111]. Measurements were taken over different time intervals in UHV conditions during ARPES measurements. (a) Only topological surface states at the Fermi level are observed at the beginning of the experiment, (b) but after $\sim$ 1h 10 min bulk conduction bands start to bend downward due to surface adsorption effect from probably the residual gases (H$_2$,CO, N$_2$). (c) Raising temperature back to room temperature and allowing the sample to recover for sometime in UHV conditions resets the amount of band bending by desorption of the adatoms. The spectrum was measured after recooling to $\sim 17$ K (d) The downward bulk band bending reappears again after $\sim$ 55 min during low-temperature ARPES measurements. These observations further support the fact that our films do not degrade in UHV conditions.
  \label{fig:ARPES_STO_111_exposure}
\end{figure*}